\documentclass[11pt]{article}
\usepackage[margin=2cm]{geometry}
\usepackage{graphicx}
\usepackage{wrapfig}
\usepackage{amsmath}

\begin{document}

\title{ODE  and  Random Boolean  networks  in  application  to  modelling of 6-mercaptopurine metabolism}
\author{Anastasia I. Lavrova$^{1,2,*}$, Eugene B. Postnikov$^3$, Andrey Yu. Zyubin$^1$, Svetlana V. Babak$^1$}

\date{$^1$ Immanuel Kant Baltic Federal University, A. Nevskogo st. 14A, Kaliningrad, Russia;\\
$^2$ St. Petersburg Research Institute of Phthisiopulmonology, \\ Polytechnicheskaya st. 32, Saint-Petersburg, Russia; \\
$^3$ Kursk State University, Radishcheva st. 33, Kursk, Russia.\\
$^*$ {\it Corresponding author, e-mail: {aurebours@googlemail.com}}
}

\maketitle

\abstract{We consider two approaches to modelling of cell metabolism of 6-mercaptopurine, which is one of the important chemotherapy drugs  used for treating of acute lymphocytic leukemia: kinetic ordinary differential equations and random Boolean networks, and analyse their interplay with respect to taking into account  ATP concentration as a key parameter of switching between different pathways.  It is shown that Boolean networks, which allows for avoiding complexity of general kinetic modelling, preserve an opportunity to the reproduction of the principal switching mechanism. To keep a detailed quantitative measure of the control parameter, a combined Boolean-ODE method is proposed. 
}

\section{Introduction}

6-mercaptopurine (6-MP) is one of the important chemotherapy drugs used for treating acute lymphocytic leukemia (ALL). It belongs to the class of medications called purine antagonists and works by stopping the growth of cancer cells. 6-MP undergoes extensive metabolic intracellular transformations that results in the production of  thionucleotides and active metabolites, which have cytotoxic and immunosuppressive  properties leading to various acute side-effects as kidney affection, hepatotoxicity, pancreatitis and neuropathy.

\begin{figure}%
\includegraphics[width=\columnwidth]{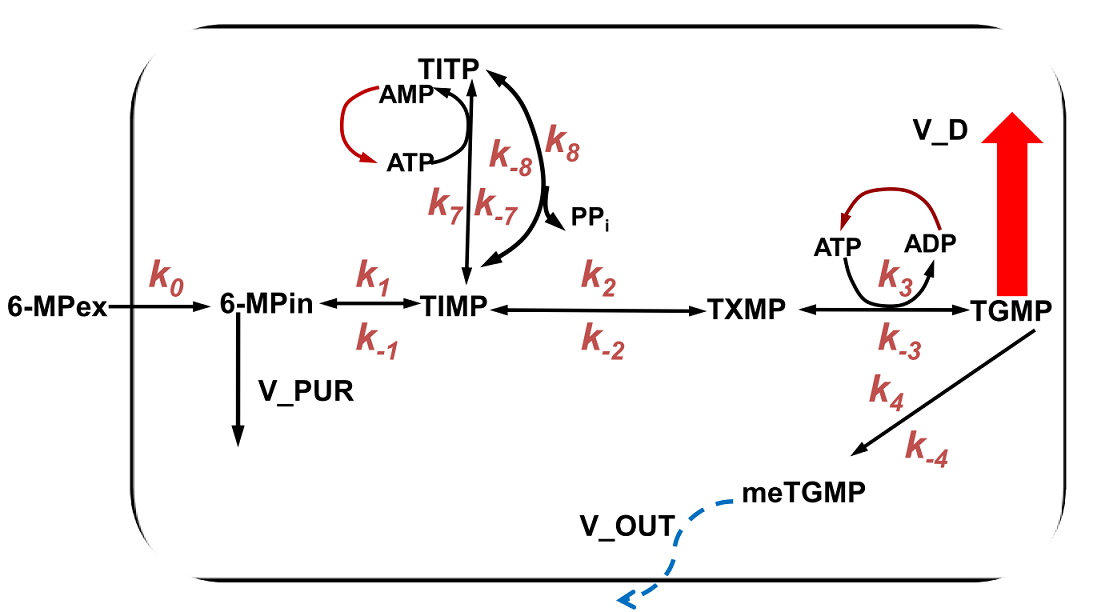}%
\caption{Simplified scheme of 6-MP metabolism. $k_i/k_{-i}$-kinetic constants of forward-back reactions; 6-MPex, 6-MP$_{in}$- mercaptopurine outside and inside of cell, TIMP, TITP- 6-Thioinosine-5'-monophosphate and triphosphate; TXMP- 6-Thioxanthine 5'-monophosphate; TGMP-
6-Thioguanosine monophosphate, meTGMP- 6-Methylthioguanosine monophosphate; ATP, ADP,AMP - adenosine tri-, di- and monophosphates; $V_D$,$V_{PUR}$,$V_{OUT}$ - common fluxes describing incorporation to DNA and RNA of cells, inhibition of purine biosynthesis {\it {de novo}} and outflux to environment}
\label{scheme}%Ы
\end{figure}

The conversion of 6-MP according the metabolic scheme presented in Fig.~\ref{scheme} involves several small pathways \cite{Cheo}. The desired pathway results in the formation of 6-Thioguanosine monophosphate (TGMP) that could be incorporated (via some metabolic transformations) into DNA and RNA leading to the tumor cell death \cite{Cheo1, Dev,Hed} in the case of successful treatment of ALL.  The catabolic pathways regulated by the enzyme mercaptopurine methyltransferase (TPMT) lead to the production of various methyl-mercaptopurines affecting purine biosynthesis \cite{Cheo1} that leads to treatment failure in most cases \cite{Dev}. The transformation (see Fig.~\ref{scheme}) of 6-Thioinosine-5'-monophosphate (TIMP) to 6-Thioinosine-5'-triphosphate (TITP) is also an additional pathway, which results in the accumulation of cytotoxic products (TITP, TDTP)  and slow production of TGMP. Since the realization of each pathway depends on enzymes properties, which are considered as the main regulators of ratio of activated and inactivated metabolites, then, in particular, a polymorphism in the corresponding genes can lead to the drug tolerance during ALL therapy \cite{Cheo1,Dor}. On the other hand, the energetic balance disturbance connected with the mitochondrial disfunction can play a crucial role in the appearance of side-effects and treatment failure \cite{Dae,FernandezRamos2016}.

Besides experimental studies, enzymes activity in 6-MP metabolism and regulation effects have been exposed to numerical simulations and mathematical modelling \cite{Dev,Cheo1,Kay}. These detailed semi-mechanistic models involve various compartments of human organism: from cell to organs to describe side-effects in dependence on the drug dose that allows for forecasting optimal dose for successful  treatment. However, these models have been more attended to the properties of regulating enzymes and exclude energy metabolism, which may play a crucial role in the occurrence of side-effects \cite{FernandezRamos2016}. Moreover, the large scale networks of interacting components require the adjusting of an enormous number of kinetic constants that prevents understanding of principal mechanisms and key parameters of switching between the pathways in 6-MP metabolism.

 This is why another approach proposed by L.~Glass and S.~Kaufmann \cite{Glass1973} gain popularity, see for a review of recent state of the art, e.g. \cite{Karlebach2008,Wang2012,LeNovere2015}: Boolean networks. The Boolean network represents a graph, whose nodes can take values 0 (inactive) or 1 (active) and the edges are matched to the rules of Boolean logic. Their evaluation with respect to the previous logical states of nodes determines the consequent state of the network's nodes. 

Certainly, if ODE-based models are over-complicated, the Boolean networks are often over-simplified. For example, sometimes their over-simplicity requires introducing tricks, which are artificial to a certain extent, like over- and under- self- expressed nodes with value $1\pm0.5$ \cite{Davidich2013}. This situation calls for some hybrid models, which should exhibit the best sides of the both approaches \cite{LeNovere2015,Fisher2007}. 

This problem is closely connected with the question about an interplay of ODEs modelling the scheme of kinetic reactions and Boolean networks simulating activity of reactants. This challenge induced a number of works, one can mention the pioneering article \cite{Davidich2008}, as well the recent developments \cite{Stoetzel2015,Menini2016}. However, the approaches considered in these works deals with the processes, which exhibit sharp transitions. In other words, such ODEs  corresponds as a rule to the high-order Hill kinetics and the extraction of fast processes is possible. It is a natural situation for the gene/protein networks, but the components kinetics of biochemical metabolic networks is more smooth.

Thus, one of the goals of the present work is an attempt to overcome this difficulty utilizing a certain freedom, which provides probabilistic Boolean networks \cite{Shmulevich2002}: a set of Boolean networks, each of them corresponds to a different pathway, and a choice between them is determined by potential interactions between underlying biological components and their uncertainties. 

\section{Kinetic ODE model}

To describe the principal dynamics of 6-MP metabolic transformations and to single out key nodes of this "metabolic chain", we have proposed a model, which describes the simplified kinetic scheme shown in Fig.~\ref{scheme}. The dimensional model does not detalize  dynamics of each enzyme but involves ATP concentration as a key player of the energy metabolism. 

As a result, the system of ODE corresponding to the simplified kinetic model can be written as follows:
\begin{align*}
\frac{d}{dt}&MP_{ex}&=&-k_0MP_{ex},\\
\frac{d}{dt}&MP_{in}&=&-(V_{PUR}+k_1)MP_{in}+k_0MP_{ex}+k_{-1}TIMP,\\
\frac{d}{dt}&TIMP&=&k_1MP_{in}+k_{-8}TITP-(k_2+k_7ATP+k_{-1}+k_8PP)TIMP\\
&&&+k_{-2}TXMP+k_{-7}TITP{\cdot}AMP,\\
\frac{d}{dt}&TXMP&=&k_2TIMP-k_3TXMP{\cdot}ATP-k_{-2}TXMP+k_{-3}TGMP{\cdot}AMP{\cdot}PP,\\
\frac{d}{dt}&TGMP&=&k_3TXMP{\cdot}ATP-(k_4+V_{D})TGMP-k_{-3}TGMP{\cdot}AMP{\cdot}PP+k_{-4}meTGMP,\\
\frac{d}{dt}&meTGMP&=&k_4TGMP-V_{OUT}meTGMP-k_{-4}meTGMP,\\
\frac{d}{dt}&TITP&=&k_8TIMP{\cdot}PP-k_{-8}TITP+k_7TIMP{\cdot}ATP-k_{-7}TITP{\cdot}AMP,\\
\frac{d}{dt}&ATP&=&-k_7TIMP{\cdot}ATP+k_{-3}TGMP{\cdot}AMP{\cdot}PP-k_3TXMP{\cdot}ATP+k_{-7}TITP{\cdot}AMP,\\
\frac{d}{dt}&AMP&=&-k_{-3}TGMP{\cdot}AMP{\cdot}PP+k_3TXMP{\cdot}ATP+k_7TIMP{\cdot}ATP-k_{-7}TITP{\cdot}AMP,\\
\frac{d}{dt}&PP&=&-k_8TIMP{\cdot}PP+k_{-8}TITP-k_{-3}TGMP{\cdot}AMP{\cdot}PP+k_3TXMP{\cdot}ATP.
\end{align*}

In our simulations, the kinetic constants corresponding to the biophysically relevant dynamics were determined as 
$k_0=5~d^{-1}$,
$k_1=10~d^{-1}$,
$k_2=10~d^{-1}$,
$k_3=5~M^{-1}d^{-1}$,
$k_4=0.00001~d^{-1}$,
$k_7=0.01~d^{-1}$,
$k_8=0.5~M^{-1}d^{-1}$,
$k_{-7}=1~M^{-1}d^{-1}$,
$k_{-1}=0.01~d^{-1}$,
$k_{-2}=4~d^{-1}$,
$k_{-3}=0.01~M^{-2}d^{-1}$,
$k_{-4}=0.1~d^{-1}$,
$k_{-8}=0.01~d^{-1}$,
$V_{PUR}=0.01~d^{-1}$,
$V_{D}=0.9~d^{-1}$,
$V_{OUT}=0.0001~d^{-1}$, 
where {\it M} means $\mu$M/mL, and d means days

The initial concentrations were equal to zero for all variables except the fixed value $MP_{ex}(0)=0.68$ $\mu$M/mL and $ATP(0)$, whose value plays a role of a control parameter.

\begin{figure}%
\includegraphics[width=0.29\textwidth]{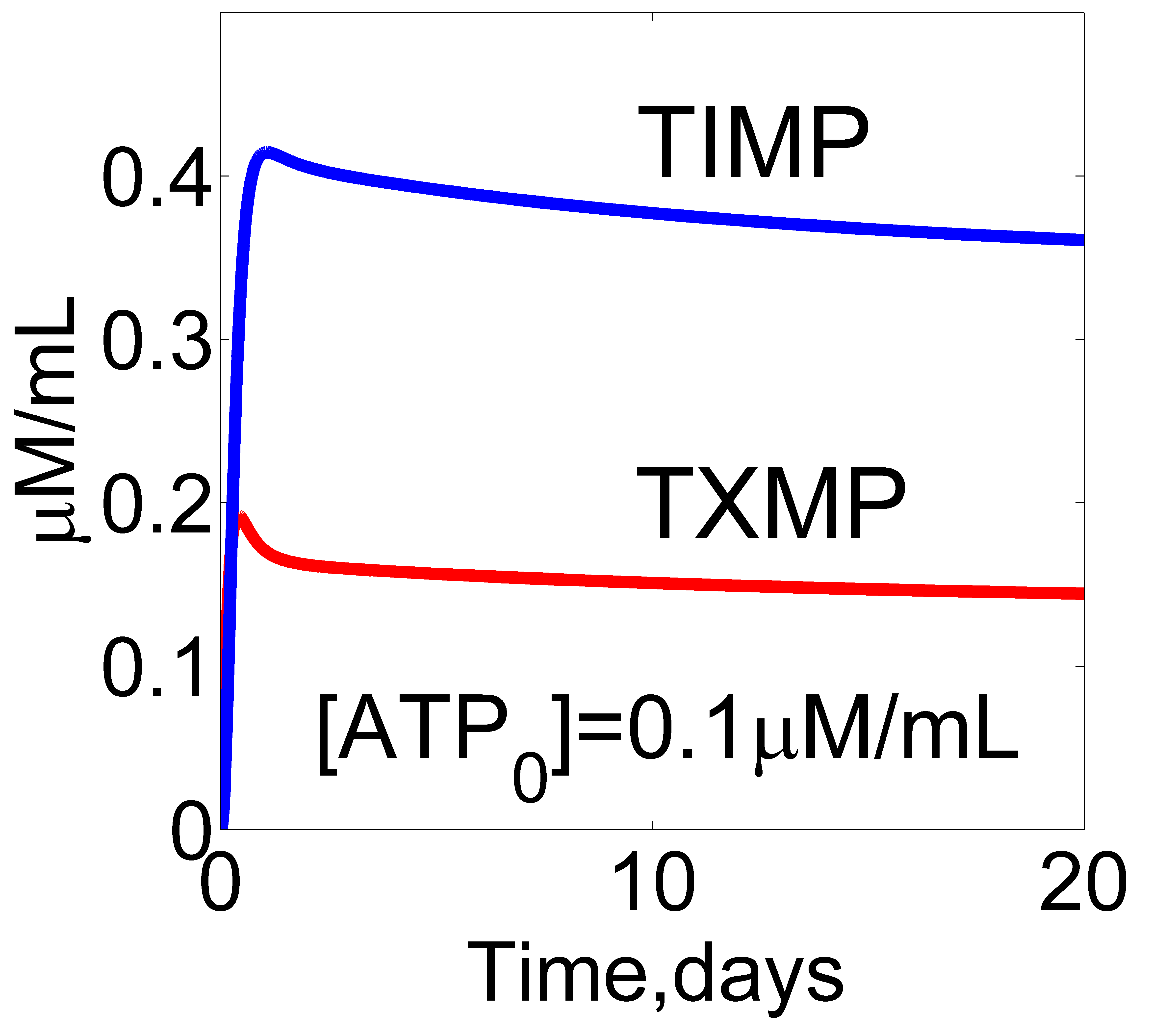}%
\includegraphics[width=0.33\textwidth]{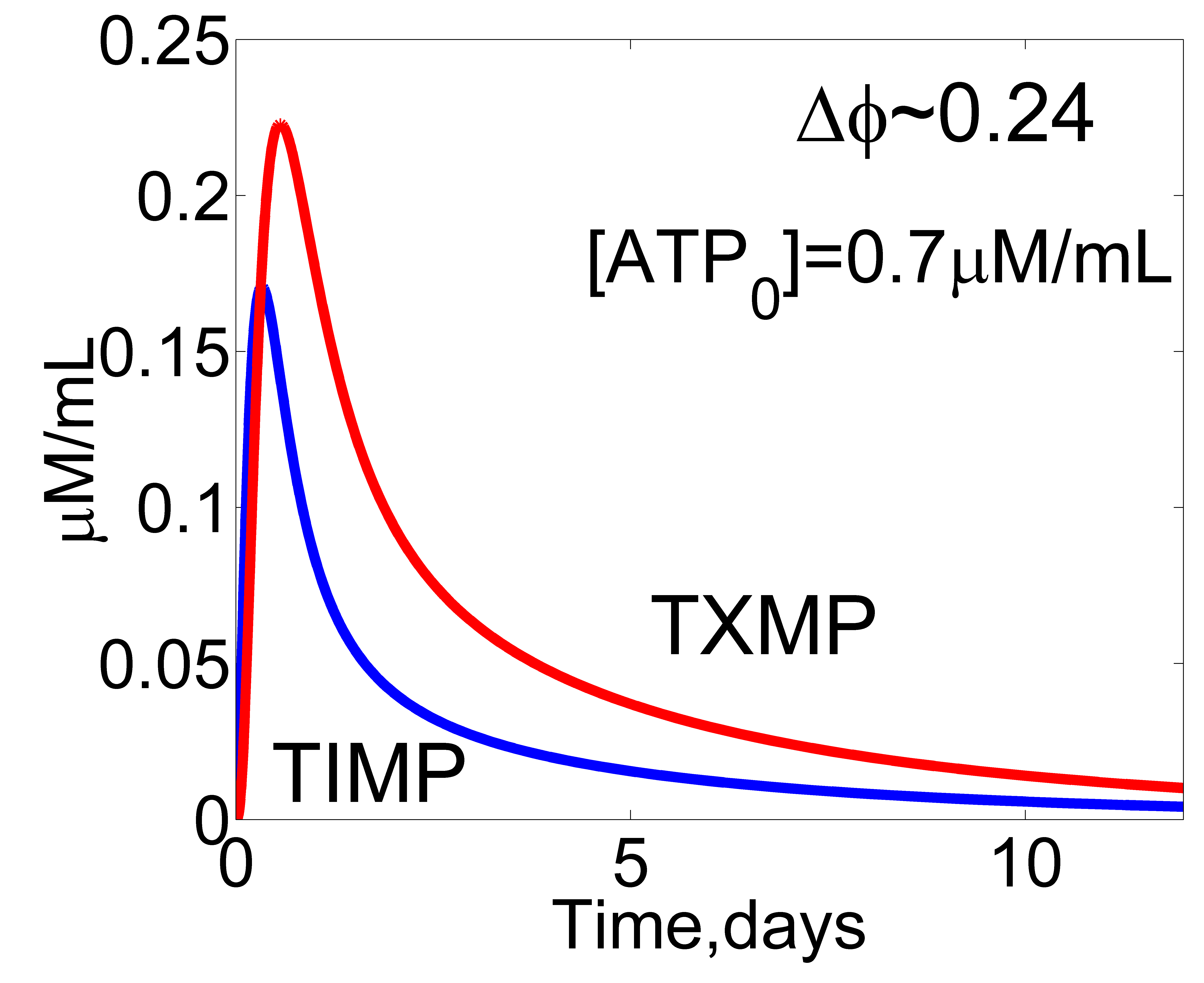}%
\includegraphics[width=0.33\textwidth]{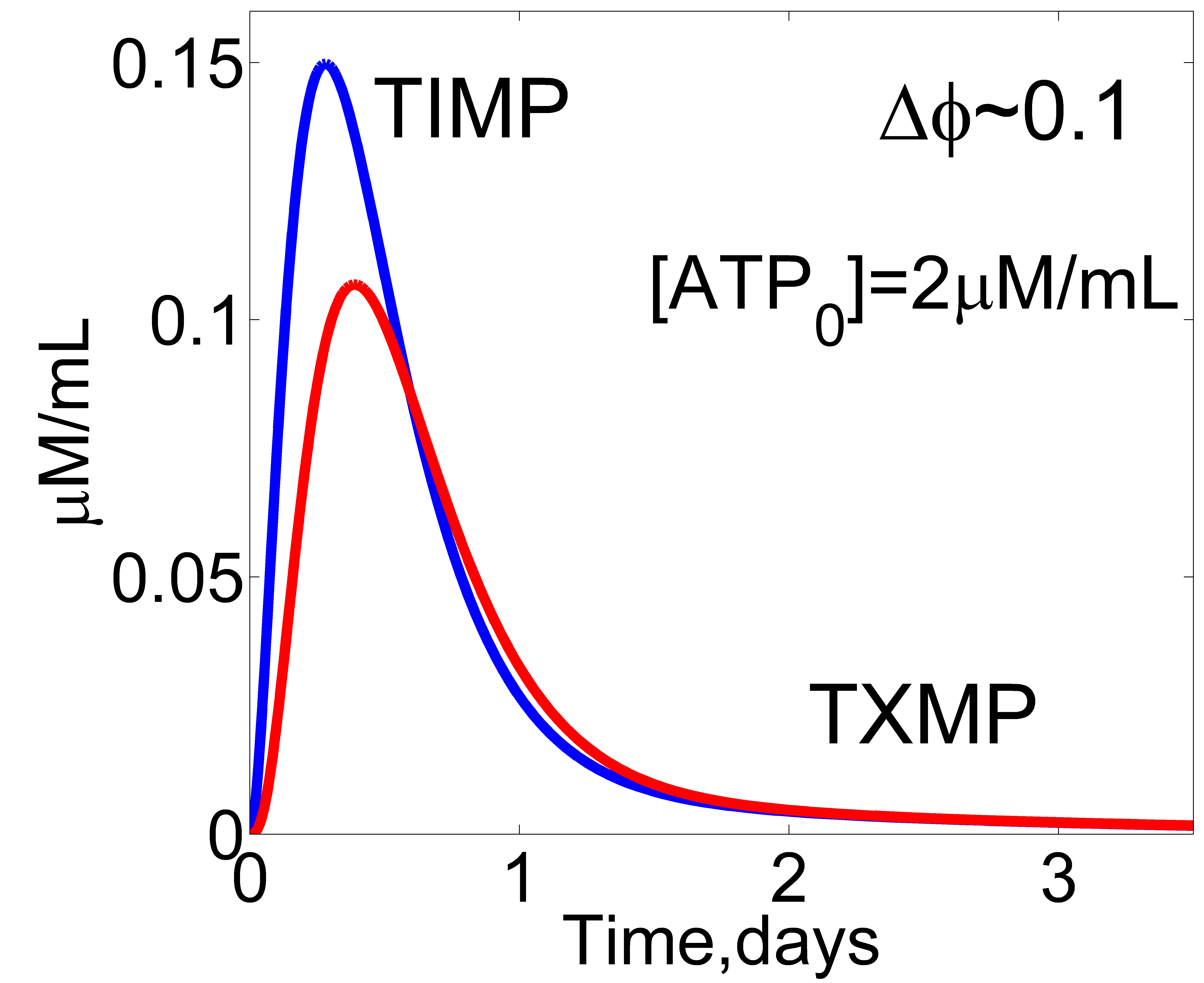}%
\caption{Dependence of the dynamics of “metabolic chain” on the initial concentration of ATP, red curve denotes TXMP concentration, blue curve is TIMP concentration}
\label{phase}%Ы
\end{figure}

\section{A Boolean network mimicking the key dynamical processes}

\subsection{Network construction}

The simplified metabolic network described above allows the representation in terms of the probabilistic Boolean network, which consists of five nodes $\{y_i\}$, $i=1..5$ and the  threshold-based rule $A(\alpha_j)$ for a choice between possible pathways. The value of the continuous control parameter $\alpha_j$ could be non-stationary in dependence of the iteration's number $j$. The correspondence of these nodes to the metabolites and the transition rules for a parallel update of states are presented in Table~\ref{booltab}.

\begin{table}[b]%
\begin{center}
\begin{tabular}{ccp{0.6\textwidth}}
\hline
Node&Metabolite&Rules of interactions and updating\\
\hline
$y_1$&6MPin&Starting node activated, when 6-mercaptopurine enters the cell. It activates TIMP and then will be deactivated.\\
$y_2$&TIMP&This node is activated by 6MPin {\it or} by TITP and can activate nodes TXMP or TITP in dependence on a chosen pathway (the choice is governed by the variable $\alpha$); it is deactivated after this.\\
$y_3$&TXMP&This node is activated by TIMP and can activate TGMP or TIMP in dependence on a chosen pathway (the choice is governed by the variable $\alpha$); it is deactivated after this.\\
$y_4$&TGMP&This node indicate the target output, is activated by TXMP and deactivated after the completed output.\\
$y_5$&TITP&This node is activated by TIMP within one of possible pathways and activate TIMP; it is deactivated after this.\\
\hline
$\alpha$&ATP&The continual parameter, which governs a choice of pathways by as follows: {\it if} $\alpha<0.5$, then then the irreversible activation TXMP$\to$TGMP is chosen; the reversible transition TXMP$\rightleftharpoons$TGMP holds otherwise; {\it if} $\alpha<0.75$ the pathway through TXMP is chosen, the pathway through TITP holds otherwise. The parameter $\alpha$ is non-stationary and satisfies the decay kinetics $\dot{\alpha}=-\kappa\alpha$ if the process goes through the TIMP pathway.\\
\hline
\end{tabular}
\end{center}
\caption{The nodes and transition rules for the considered Boolean network}
\label{booltab}
\end{table}

The realisation of this its  via Boolean and conditional operators reads as follows (here the states of nodes are grouped into the matrix $y(j,i)$):
\begin{verbatim}
y(:,1)=[1 0 0 0 0]';
for j=2:M;
    if y(1,j-1)==1; 
       y(2,j)=1; 
    end
    if alpha>0.5  
       if (y(2,j-1)==1)|(y(5,j-1)==1)    
          if alpha<0.75
              y(3,j)=y(2,j-1);
              y(2,j)=y(5,j-1);
          else
              y(5,j)=1;
              alpha=(1-kappa)*alpha;
          end
        end  
            y(4,j)=y(3,j-1);
    else
        y(3,j)=y(2,j-1);
        y(2,j)=1;
    end; 
end 
\end{verbatim}

Here ``\verb!=!'', ``\verb!==!'', and ``\verb!|!'' operators denotes the assignment, the equality, and OR respectively. Note that the code represented above can be evaluated straightforwardly using MATLAB or other software, which support MATLAB-like syntax (e.g. OCTAVE, FreeMat) if supplied by initial conditions and a value of the decay parameter. The last one is introduced via the simplest discretization of of the equation $\dot{\alpha}=-\kappa\alpha$ via the Euler scheme with the unit time step (i.e. in accordance to the assumed step of networks nodes updating):
$\alpha_{j+1}=(1-\kappa)\alpha_j$. 

For example, they may be stated as
\begin{verbatim}
N=5; % Number of nodes
M=8; % Number of iterations
y=zeros(N,M); % Initializing the matrix
ATP=0.6; % Initial ATP input (control parameter)
kappa=0.1; % The decay parameter value
alpha=ATP;
\end{verbatim}

\subsection{Simulation results}

The growth of  metabolites concentrations occurs sequentially for one node after another along the “metabolic chain” (see Fig.~\ref{scheme}).  It is revealed that TIMP is the key node of the  reactions cascade  since it provides two pathways, slow and fast,  that also defines the blockage of the slow way interacting with ATP. Here we can define the concentration of ATP as a “key player” in the 6-MP metabolism, which regulates transitions in two main points: the metabolic pathway  of TITP production (the chain’s branch from TIMP) and  the transition TXMP$\to$TGMP. 

Simulations of this kinetic model show that small concentrations of ATP lead to the blockage of metabolic chain in the node TIMP (Fig ~\ref{phase},left). Large concentrations  of ATP result in the competition between production of TITP (the end product of 
branch) and  TGMP (the product of the chain), see (Fig ~\ref{phase},middle). The optimal concentration of ATP, which shifts the pathway to higher production of TGMP is equal to 0.7 $\mu$M/mL ((Fig ~\ref{phase},right)).

The same situation can be observed at the simulation of Boolean Networks. Table~\ref{statestab} represents the results of simulations evaluated for a set of increasing initial values (\verb!ATP!) of the control parameter $\alpha$. They capture all principal features of the dynamics for the simulated network. Note that first two steps are the same for all cases since the activation of TIMP by 6-MP$_{in}$ is unconditional. The different pathways are realized during next iterations only. 

For \verb!ATP=0.2! the dynamics is blocked at the transition from TXMP to TGMP. Instead of the forward activation, the process goes back reactivating the node corresponding to TIMP. At the same time, since this reaction is reversible, the reactivation of TXMP occurs, etc. Thus, the system reaches a steady state, which is reflected in unit values of the nodes $y_2$ and $y_3$ spreading {\it ad infinitum}. 

The value \verb!ATP=0.2! corresponds to the situation, where the pathway TXMP $\to$ TGMP is allowed but the pathway leading to TITP is blocked. As a result, the transition process is direct and straightforward: the nodes $y_1$--$y_4$ are activated sequentially during the four sequential iterations. When $y_4$ is activated, this means that the target substance is released, and all nodes switch off to zeros in absence of a new influx into $y_1$.  

Both the values \verb!ATP=0.8! and \verb!ATP=0.9! exceed the threshold value $\alpha=0.75$. Whence, the pathway to TITP is available now. It is reflected as $y_3=0$ but $y_5=1$ at the third iteration, i.e. the pathway is changed. However, there is a difference in the further time evolution of the network's states for these two cases. Namely, the table corresponding to 
 \verb!ATP=0.8! demonstrate the activation of $y_3$ (i.e. the backward transition TITP$\to$ TIMP) during the next iteration and the consequent sequential nodes activation along the pathway TIMP$\to$ TXMP$\to$ TGMP. On the other hand, these step are delayed in the case of \verb!ATP=0.8!: the both 3rd and 4th iterations contains $y_5=1$ and $y_i=0$, $i=1..4$ only. 
Such a behaviour originates from the introduced non-stationarity of the control parameter $\alpha$, which resembles the concentration of ATP. As it was discussed above, the TIMP$\to$ TITP pathway is an ATP-consuming process. Thus, each iteration corresponding to this pathway diminishes $\alpha$ while it will cross the threshold $\alpha=0.75$ from above. Further, this pathway will be blocked. The cases \verb!ATP=0.8! and \verb!ATP=0.9! require one and two iterations for this decay of $\alpha$, respectively. Larger values of \verb!ATP! will result in larger delays. 

Finally, we should note that the discussed results are quasi-deterministic, since they 
correspond to individual realizations. In a general case, i.e. in the strict sense of probabilistic Boolean networks, one can generate an ensemble of realisations with \verb!ATP! randomly distributed with respect to some appropriate probability distribution. Correspondingly, the output will be a distribution of the nodes values during iterations. But this procedure is out of direct goals of the present work. 

\begin{table}[h]%
\begin{center}
\begin{tabular}{c|ccccccc}
ATP=0.2&&&&&&&\\
\hline
$j$& 1& 2& 3& 4& 5& 6& 7\\
\hline
$y_1$& 1&	0&	0&	0&	0&	0&	0\\
$y_2$& 0&	1&	1&	1&	1&	1&	1\\
$y_3$& 0&	0&	1&	1&	1&	1&	1\\
$y_4$& 0&	0&	0&	0&	0&	0&	0\\
$y_5$& 0&	0&	0&	0&	0&	0&	0\\
\end{tabular}

\begin{tabular}{c|ccccccc}
ATP=0.6&&&&&&&\\
\hline
$j$& 1& 2& 3& 4& 5& 6& 7\\
\hline
$y_1$& 1&	0&	0&	0&	0&	0&	0\\
$y_2$& 0&	1&	0&	0&	0&	0&	0\\
$y_3$& 0&	0&	1&	0&	0&	0&	0\\
$y_4$& 0&	0&	0&	1&	0&	0&	0\\
$y_5$& 0&	0&	0&	0&	0&	0&	0\\
\end{tabular}

\begin{tabular}{c|ccccccc}
ATP=0.8&&&&&&&\\
\hline
$j$& 1& 2& 3& 4& 5& 6& 7\\
\hline
$y_1$& 1&	0&	0&	0&	0&	0&	0\\
$y_2$& 0&	1&	0&	1&	0&	0&	0\\
$y_3$& 0&	0&	0&	0&	1&	0&	0\\
$y_4$& 0&	0&	0&	0&	0&	1&	0\\
$y_5$& 0&	0&	1&	0&	0&	0&	0\\
\end{tabular}

\begin{tabular}{c|ccccccc}
ATP=0.9&&&&&&&\\
\hline
$j$& 1& 2& 3& 4& 5& 6& 7\\
\hline
$y_1$& 1&	0&	0&	0&	0&	0&	0\\
$y_2$& 0&	1&	0&	0&	1&	0&	0\\
$y_3$& 0&	0&	0&	0&	0&	1&	0\\
$y_4$& 0&	0&	0&	0&	0&	0&	1\\
$y_5$& 0&	0&	1&	1&	0&	0&	0\\
\end{tabular}
\end{center}
\caption{The evolution of network states for different various values of the control parameter.}
\label{statestab}
\end{table}

\section{Discussion}

It is known that the  methylation of  6-MP resulted in a formation of the intermediate metabolites occurs at a low concentration of intracellular ATP (0.1~$\mu$mol/ml).
Simultaneously, the concentrations  TIMP and TXMP remain at a prolonged constant level. At these conditions, a production of the final metabolite, TGMP, slows down. As a result, the therapeutic efficiency also diminishes but the risk of toxic action grows since intermediate metabolites of 6-MP inhibit biosynthesis of {\it de novo} purines. Thus, the lowering intracellular ATP pool in T-lymphocytes results in higher toxicity and lower efficiency of this drug \cite{FernandezRamos2016,Valente2016}.  

Our results model an effect of high initial concentration of ATP on the metabolism of 6-MP. They show that ATP concentrations 0.1~$\mu$mol/ml produce high concentrations of the intermediate metabolite $TIMP$ that indicates an incomplete metabolism of the drug accompanied by a production of TGMP insufficient for the therapeutic action. Therefore, we suppose that the intensive TIMP formation plays a role of the marker indicating an accumulation of toxic final metabolites at a high level of intracellular ATP.

A concentration change of ATP is a key factor for the energy exchange deficit  accompanied by the mitochondrial \cite{Beuster2011}. This results in decreasing therapeutic effect of drugs during the tumor treatment. It has been shown the glycolysis inhibition by an attenuation of the glucose consumption cell function leads to the diminishing of ATP level and, finally, results in tumor cell death. However, the process of energy deficiency is invertible since the cell activates another pathway, which supports ATP accumulation, and the cell will recover its function.

The results obtained using our model argue that the optimal initial ATP concentration is equal to 0.7~$\mu$mol/ml. It corresponds to the situation, when 6-MP metabolism is a completed process resulted in the both production of therapeutically active products and reducing the pool of toxic intermediate products.

At the administration of cytotoxic drugs according to the protocol BFM ALL 2000 \cite{Flohr2008}, it is expedient to keep the concentration of intracellular ATP within a middle range to prevent risks of adverse drugs reactions instead of an artificial inhibition of energy metabolism \cite{Beuster2011}. 

The clinical indication of low ATP concentration is acidosis by lactates accumulation \cite{Beuster2011}. This leads to the mitochondrial dysfunction and an additional toxic effect. Higher ATP concentrations inhibit glycolysis resulting in a glucose accumulation, glycose tolerance, and, indirectly, in the cardiomyopathy development \cite{Guertl2011}.    

Thus, we hypothesize that the maintenance of ATP intermediate level is a necessary condition to reach a complete therapeutic effect and diminish toxicity of a chemotherapy process.

\section{Conclusion and Outlooks}

In this work, we have analysed the dynamic behaviour of metabolic pathways of 6-mercaptopurine with a focus on the revealing a key parameter, which switches between two principal ``branches'', slow and fast one. The results of simulations based on the system of ordinary equations indicate that ATP is the desired “key player” in the 6-MP metabolism. This conclusion is supported by a number of phenomenological observations presented in the modern biomedical literature and allows for quantitative clarifying the underlying processes.

Basing on the results of ODE modelling, we have reformulated the problems in terms of the probabilistic Boolean networks. This approach is much more simpler in realization since it does not require a knowledge of multiple kinetic parameters but, in the same time, adequately reproduces the key details of switching principal dynamic regimes as a choice between different possible pathways. Therefore, it can be scaled to more detailed picture of metabolites interactions in future research of the studied process. 

We also need to highlight the crucial feature introduced into a construction of the network: a non-stationary continual parameter, which governs the switching process. Such an approach, which has demonstrated its effectiveness in the considered case study, opens new perspectives for ``hybridising'' of continual (ODE-based) and discrete (Boolean) approaches to metabolic modelling. In contrast to previous works \cite{Davidich2008,Stoetzel2015,Menini2016}, which considered Boolean networks only as a limiting case of continual-time kinetic processes (in fact, as a mimicking of switching between unstable stationary state by nodes activity), the introduction of non-stationarity into the probabilistic parameter allows the consideration of smoother transitions, and, in principle, even an activity of small sub-networks with a small number of kinetic constants considered as building blocks for a large Boolean network. 

\section*{Acknowledgement}

The work is supported by the Grant no. 14.575.21.0073, code RFMEFI57514X0073 of the Ministry of Education and Science of the
Russian Federation

\end{document}